\documentclass[preprint]{revtex4}

\usepackage{graphicx}
\usepackage{mathrsfs}
\usepackage{amsmath}

\def\iso#1#2{\mbox{${}^{#2}{\rm #1}$}}

\def\c1#1{\iso{C}{1#1}}
\def\n1#1{\iso{N}{1#1}}
\def\o1#1{\iso{O}{1#1}}

\newcommand{\be}{\begin{equation}}
\newcommand{\ee}{\end{equation}}

\def\beq#1\eeq{\begin{equation}#1\end{equation}}
\def\beqar#1\eeqar{\begin{eqnarray}#1\end{eqnarray}}

\def\la{\mathrel{\mathpalette\fun <}}
\def\ga{\mathrel{\mathpalette\fun >}}
\def\fun#1#2{\lower3.6pt\vbox{\baselineskip0pt\lineskip.9pt
  \ialign{$\mathsurround=0pt#1\hfil##\hfil$\crcr#2\crcr\sim\crcr}}}
\def\ie{{\it i.e.},~}
\def\eg{{\it e.g.},~}
\def\etal{{\it et al.}}
\def\Yp{Y$_{\rm P}$}

\def\hii{H\thinspace{$\scriptstyle{\rm II}$}}

\def\4he{$^4$He}
\def\3he{$^3$He}
\def\4he{$^4$He}
\def\7li{$^7$Li}
\newcommand{\Deln}{\ensuremath{\Delta {\rm N}_\nu}}
\newcommand{\nnu}{\ensuremath{{\rm N}_\nu}}
\newcommand{\epm}{\ensuremath{e^{\pm}\;}}

\begin{document}

\title {Primordial Helium And the Cosmic Background Radiation}

\author{Gary Steigman}
\email{steigman@mps.ohio-state.edu}
\affiliation{Departments of Physics and Astronomy, Ohio State University \\
191 W. Woodruff Ave., Columbus OH 43210-1117, USA}

\begin{abstract}
The products of primordial nucleosynthesis, along with the 
cosmic microwave background (CMB) photons, are relics from 
the early evolution of the Universe whose observations probe 
the standard model of cosmology and provide windows on new 
physics beyond the standard models of cosmology and of particle 
physics.  According to the standard, hot big bang cosmology, 
long before any stars have formed a significant fraction ($\sim 
25\%$) of the baryonic mass in the Universe should be in the 
form of helium-4 nuclei.  Since current observations of \4he are 
restricted to low redshift regions where stellar nucleosynthesis 
has occurred, an observation of high redshift, prestellar, truly 
primordial \4he would constitute a fundamental test of the hot, 
big bang cosmology.  At recombination, long after big bang 
nucleosynthesis (BBN) has ended, the temperature anisotropy 
spectrum imprinted on the CMB depends on the \4he abundance 
through its connection to the electron density and the effect of 
the electron density on Silk damping.  Since the relic abundance 
of \4he is relatively insensitive to the universal density of baryons,
but is sensitive to a non-standard, early Universe expansion rate, 
the primordial mass fraction of \4he, \Yp, offers a test of the 
consistency of the standard models of BBN and the CMB and, 
provides constraints on non-standard physics.  Here, the WMAP 
seven year data (supplemented by other CMB experiments), which 
lead to an indirect determination of \Yp~at high redshift, are 
compared to the BBN predictions and to the independent, direct 
observations of \4he in low redshift, extragalactic \hii~regions.  
At present, given the very large uncertainties in the CMB-determined 
primordial \4he abundance (as well as for the helium abundances 
inferred from \hii~region observations), any differences between 
the BBN predictions and the CMB observations are small, at a 
level $\la 1.5\sigma$.

\end{abstract}

\maketitle

\section{Introduction}

In the first few minutes of the evolution of the Universe, during big 
bang nucleosynthesis (BBN), neutrons and protons are incorporated 
in astrophysically interesting (\ie measurable) abundances into the 
light nuclides D, \3he, \4he, and \7li.  Of these nuclides, the relic 
abundance of \4he (mass fraction \Yp) is least sensitive to the nuclear 
reaction rates (and their uncertainties) and to the baryon density but, 
\Yp~is very sensitive to the early universe energy density through its 
effect on the expansion rate of the Universe during radiation-dominated 
epochs.  As a result, a comparison of the BBN-predicted value of 
\Yp~with its observationally determined value has the potential to 
test the standard models of particle physics and cosmology and to 
constrain any new physics beyond the standard models (see, \eg 
\citet{ssg,boes,steigman07}, and references therein).  For several 
decades observations of \4he (and H) recombination lines in 
metal-poor, extragalactic \hii~regions have provided the data 
needed to infer the primordial \4he mass fraction which may be 
compared with the predictions of BBN in the standard model (SBBN).  
Over the years, the observationally inferred \4he abundance has 
varied (largely increasing) from \Yp~$\la 0.23$ to \Yp~$\ga 0.25$, 
the variations due, in large part, to better data and to better 
analyses of the data which, in particular, address the systematic 
uncertainties in using the observed hydrogen and helium recombination 
line intensities to derive the \4he abundances.  Cosmological 
tests require that \Yp~be known to $\la 0.4\%$.  An 
{\bf independent} determination of \Yp, with different systematics, 
would be of great value.  Furthermore, the \hii~region observations 
of \4he are in low redshift, star-forming regions, polluted to 
some extent by the products of stellar nucleosynthesis.  The CMB 
can play an important role by providing a completely independent 
determination of \Yp~in the high redshift, post-BBN, prestellar 
Universe, free from the systematic uncertainties affecting the 
direct observations of \4he in extragalactic \hii~regions.
 
\section{Standard BBN And Primordial Deuterium}
\label{SBBND}

Long before recombination, during the first few minutes in the evolution
of the Universe, the light nuclides D, \3he, \4he, and \7li are synthesized
by BBN.  For ``standard" BBN (SBBN), with three flavors of light neutrinos 
(\nnu~= 3), the light element abundances depend on only one adjustable 
parameter, the baryon (or nucleon) abundance, $\eta_{\rm B} \equiv 
n_{\rm B}/n_{\gamma} \equiv 10^{-10}\eta_{10}$.  In the standard models 
of particle physics and cosmology, the numbers of nucleons and CMB
photons in every comoving volume are preserved (post-\epm annihilation),
so that $\eta_{\rm B}$ should be unchanged from BBN to recombination 
to the present. Of the relic nuclides, deuterium is the baryometer of choice 
since its post-BBN evolution is simple and monotonic.  As gas is cycled 
through stars, deuterium is destroyed and no significant amounts of D 
are synthesized in stellar (or other) post-BBN nucleosynthesis~\cite{els}.  
The abundance of deuterium, observed anywhere in the Universe, at any 
time in its post-BBN evolution, is never any larger than the primordial 
D abundance.  Furthermore, the BBN-predicted D abundance is sensitive 
to the baryon density parameter, varying (for $\eta_{\rm B}$ in the range 
of interest) as $\eta_{\rm B}^{-1.6}$, so that a $\sim 10\%$ determination 
of $y_{\rm DP} \equiv 10^{5}$(D/H)$_{\rm P}$ results in a $\sim 6\%$ 
measurement of the universal density of baryons.

Nearly primordial deuterium is seen in absorption against background UV 
sources in QSO Absorption Lines Systems (QSOALS).  These observations are 
difficult, requiring significant time on large telescopes equipped with 
high resolution spectrographs.  At present there are determinations of the 
deuterium abundance along only seven, high-redshift, low-metallicity lines 
of sight.  The results, summarized in \citet{pettini}, lead to a primordial 
D abundance log~$y_{\rm DP} = 0.45 \pm 0.03$.  For SBBN (\nnu~= 3), this 
corresponds to a baryon abundance $\eta_{10}($SBBN,D$) = 5.80^{+0.27}_{-0.28}$ 
($\Omega_{\rm B}h^{2} = 0.0212 \pm 0.0010$).  For this baryon abundance, 
the SBBN-predicted primordial \4he mass fraction is \Yp(SBBN,D)~$= 0.2482 
\pm 0.0007$~\cite{ks,steigman07}.

\section{Standard BBN And The WMAP Baryon Density}
\label{SBBNW}

The CMB temperature anisotropy spectrum probes the baryon abundance 
(among many other cosmological parameters) at recombination.  Using 
the WMAP Seven-Year data, \citet{wmap7} derive $\eta_{10}($WMAP$) = 
6.190 \pm 0.145$ ($\Omega_{\rm B}h^{2} = 0.02260 \pm 0.00053$).  
This determination of the baryon abundance at recombination, some 
$\sim$ 400 kyr after BBN, provides an independent determination of 
$\eta_{\rm B}$.  Although somewhat higher than the value found from 
SBBN and deuterium (see \S \ref{SBBND}), the two determinations differ 
by only $\sim 1.3\sigma$: $\eta_{10}($SBBN,WMAP$) - \eta_{10}($SBBN,D$) 
= 0.39^{+0.32}_{-0.31}$.  For SBBN with this baryon abundance, the 
primordial deuterium abundance is predicted to be log~$y_{\rm 
DP}($SBBN,WMAP$) = 0.405^{+0.020}_{-0.021}$, which differs from 
the observationally determined value~\cite{pettini} by only $\sim$ 
1.1$\sigma$.  The SBBN/WMAP-predicted primordial \4he mass fraction 
is \Yp(SBBN,WMAP) $= 0.2488 \pm 0.0006$ \cite{ks,steigman07}.  
Whether SBBN and deuterium or SBBN and WMAP is used to find \Yp, 
the SBBN-predicted relic abundance is \Yp~$\approx 0.248 - 0.249$, 
with a $\sim$ $0.2 - 0.3\%$ uncertainty.

\section{Non-Standard BBN And WMAP}
\label{nsbbn}

A non-standard early Universe expansion rate (Hubble parameter: 
$H' \neq H$; $S \equiv H'/H \neq 1$) characterizes a large class of 
non-standard cosmological and particle physics models.  During the 
early evolution of the Universe, a non-standard expansion rate may 
be expressed in terms of an expansion rate parameter, $S$, as $S 
\equiv H'/H = (G'\rho'/G\rho)^{1/2}$, where $G$ is the gravitational 
constant and $\rho$ is the energy density, dominated at early epochs 
by the contribution from massless or extremely relativistic particles 
(``radiation").  One possibility is a non-standard energy density: 
$\rho \rightarrow \rho'  \equiv \rho + \Delta$N$_{\nu}\rho_{\nu}$, 
where the non-standard contribution is normalized to $\rho_{\nu}$, 
the total energy density from one, two-component, relativistic neutrino.  
The parameter \Deln~$\equiv$~N$_{\nu} - 3$, the ``effective number of 
equivalent neutrinos", is a convenient way to characterize a non-standard 
(\Deln~$\neq 0$), early Universe expansion rate, but it need not actually 
count new flavors of neutrinos~\cite{ssg,ks,steigman07}.  In the standard 
model (\nnu~= 3), at $T \ga m_{e}$, prior to \epm annihilation, $T_{e} 
= T_{\nu} = T_{\gamma}$, so that $\rho_{e}/\rho_{\gamma} = 7/4$, 
$\rho_{\nu}/\rho_{\gamma} = 7/8$, and
\beq
\rho = \rho_{\gamma} + \rho_{e} + 3\rho_{\nu} = 43\rho_{\gamma}/8.
\eeq
For a non-standard cosmology, $S(\neq 1$) and \Deln($\neq 0$) 
are related by $S = (1 + 7\Delta$N$_{\nu}/43)^{1/2}$~\cite{ks}.

The effect on BBN of a non-standard expansion rate is to modify the 
competition between the nuclear (and weak) reaction rates and the 
universal expansion rate.  Since the primordial \4he mass fraction 
is largely determined by the neutron to proton ratio when BBN begins 
in earnest, \Yp~is quite sensitive to the competition between the 
weak interaction rates (\ie $\beta$-decay) and the expansion 
rate~\cite{ssg}.  For $|\Delta$N$_{\nu}| \la 1$, $\Delta$Y$_{\rm P} 
\approx 0.013\Delta$N$_{\nu}$.  There are relatively smaller, but 
non-negligible changes to the BBN-predicted abundances of the 
other light elements; see, \eg \citet{ks} and~\citet{steigman07}.

When \epm pairs annihilate ($T \la m_{e}$), the photons are heated with 
respect to the neutrinos.  On the quite good assumption that the e-, 
$\mu$-, and $\tau$-neutrinos are decoupled (from the photon-\epm plasma) 
when $T \approx m_{e}$, then after \epm annihilation $T_{\gamma}/T_{\nu} 
= (11/4)^{1/3}$.  In the radiation-dominated, post-\epm annihilation 
Universe the energy density in the standard model is,
\beq
\rho = \rho_{\gamma} + 3\rho_{\nu} = [1 + (21/8)(4/11)^{4/3}]\rho_{\gamma} 
= 1.68\rho_{\gamma}.
\eeq
For a non-standard model, in the approximation of complete neutrino 
decoupling, $\rho' = [1 + (7/8)(4/11)^{4/3}$N$_{\nu}]\rho_{\gamma}$, 
where \nnu~= 3 + \Deln.  A non-standard energy density or expansion rate 
affects the transition from radiation domination to matter domination, 
impacting the growth of perturbations, and leaving an imprint on the 
CMB temperature anisotropy spectrum.  As a result, the CMB provides 
a probe of \nnu~which is independent of BBN..

However, since the e-, $\mu$-, and $\tau$-neutrinos are {\bf not} fully 
decoupled at \epm annihilation, the effective number of equivalent 
neutrinos at recombination is not \nnu, but \nnu~$\rightarrow$~N$_{\rm 
eff} = 3.046$ + \Deln~\cite{mangano}.  Since~\citet{wmap7} adopt N$_{\rm 
eff} \equiv 3.04$ + \Deln, in the comparison here with the WMAP 
seven-year data, \Deln~$\equiv$~N$_{\rm eff} - 3.04$ will be used.  

The CMB temperature anisotropy spectrum determines $z_{\rm eq}$, the 
redshift of the epoch of equal radiation and matter densities: $1 + 
z_{\rm eq} = \Omega_{\rm M}/\Omega_{\rm R}$.  Since $\Omega_{\rm R}$ 
depends on N$_{\rm eff}$, a CMB determination of N$_{\rm eff}$ is 
degenerate with the energy density in non-relativistic matter 
($\Omega_{\rm M}h^{2}$)~\cite{wmap7}.  According to~\citet{wmap7},
\beq
{\rm N}_{\rm eff} - 3.04 = 7.44\Biggl({\Omega_{\rm M}h^{2} 
\over 0.1308} {3139 \over 1 + z_{\rm eq}} - 1\Biggr) \equiv \Delta{\rm N}_{\nu}.
\eeq
To constrain N$_{\rm eff}$, the CMB data needs to be supplemented by 
independent, external data on the Hubble parameter ($H_{0}$) and on 
$\Omega_{\rm M}$ (or, $\Omega_{\rm M}h$) from observations of large 
scale structure (LSS).  For their constraint on N$_{\rm eff}$, 
\citet{wmap7} adopt the improved measurement of $H_{0}$ from \citet{riess} 
and LSS data either from baryon acoustic oscillations (WMAP+BAO+$H_{0}$) 
or from luminous red galaxies (WMAP+LRG+$H_{0}$).

For WMAP+BAO+$H_{0}$,~\citet{wmap7} find N$_{\rm eff}=4.34^{+0.86}_{-0.88}$,
corresponding to \Deln(WMAP+BAO+$H_{0}$) $= 1.30^{+0.86}_{-0.88}$, while 
for WMAP+LRG+$H_{0}$, they find N$_{\rm eff}=4.25^{+0.76}_{-0.80}$, 
corresponding to \Deln(WMAP+LRG+$H_{0}$) $= 1.21^{+0.76}_{-0.80}$.  At 
$\sim1.5\sigma$, these CMB/LSS results are consistent with the standard 
model value of \Deln~= 0 (\nnu~= 3).

For the WMAP value of  the baryon density parameter ($\eta_{10}({\rm WMAP}) 
= 6.190 \pm 0.145$) and either the WMAP+BAO+$H_{0}$ or WMAP+LRG+$H_{0}$ 
determinations of \Deln~\cite{wmap7}, the BBN-predicted deuterium abundance 
\cite{ks,steigman07} is log~$y_{\rm DP} = 0.47 \pm 0.05$, in excellent 
agreement with the observationally-determined value~\cite{pettini}, 
log~$y_{\rm DP} = 0.45 \pm 0.03$.  The BBN-predicted \4he mass 
fractions are \Yp(WMAP+BAO+$H_{0}$)~$= 0.2649^{+0.0099}_{-0.0108}$ 
and \Yp(WMAP+LRG+$H_{0}$)~$= 0.2639^{+0.0088}_{-0.0098}$, respectively.  
Although these values of \Yp~may seem high, the uncertainties are large 
(reflecting the large uncertainties in the WMAP determination of \Deln) 
and these abundances are consistent with those for SBBN (see \S\ref{SBBND} 
\& \S\ref{SBBNW}) within $\sim 1.5\sigma$.

\section{Primordial Helium-4 From The CMB}
\label{cmb}

The suppression of the CMB temperature power spectrum on small angular 
scales due to Silk damping \cite{silk} provides an independent probe of 
the relic, prestellar \4he abundance, through its effect on the electron 
density at recombination.  After helium recombination, but prior to 
hydrogen recombination, the number density of free electrons (which are 
responsible for Silk damping) is related to the baryon number density 
by $n_{e} = (1 - $Y$_{\rm P})n_{\rm B} \propto (1 - $Y$_{\rm P})\eta_{\rm 
B}$.  The  larger \Yp, the fewer free electrons, the further can the CMB 
photons free-stream, damping perturbations in the temperature anisotropy 
spectrum, reducing the CMB power spectrum on small angular scales.  Since 
this effect is largest on the smallest angular scales, the WMAP data needs 
to be supplemented by data from small-scale CMB experiments such as 
ACBAR~\cite{acbar} and QUaD~\cite{quad}.  For WMAP data alone, 
\citet{wmap7} find a 95\% upper limit to the primordial helium mass fraction of 
\Yp(CMB)~$< 0.51$.  When ACBAR and QUaD data are added, \citet{wmap7} 
find \Yp(CMB)~$= 0.326 \pm 0.075$, a result which differs from zero at more 
than $3\sigma$ (but, apparently, not by the $5\sigma$ usually required to 
establish new discoveries).  

Although the central value of this determination of \Yp~seems high, its 
uncertainty is large.  For example, it is interesting to test the internal 
consistency of this independent, CMB determination by comparing it 
to the WMAP+BAO+$H_{0}$, BBN-predicted value 
(\S\ref{nsbbn}).
\beq
{\rm Y}_{\rm P}({\rm CMB}) - {\rm Y}_{\rm P}({\rm WMAP+BAO+}H_{0}) = 0.061
\pm 0.76.
\eeq
This result is consistent with zero at $\sim 0.8\sigma$ (as is that using 
WMAP+LRG+$H_{0}$).  It is also interesting to compare the CMB result to 
the one determined by SBBN and the observed D abundance (\S\ref{SBBND}),
\beq
{\rm Y}_{\rm P}({\rm CMB}) - {\rm Y}_{\rm P}({\rm SBBN,D}) = 0.078 \pm 0.075,
\eeq
or, with that predicted by SBBN using the WMAP-determined baryon abundance 
(\S\ref{SBBNW}),
\beq
{\rm Y}_{\rm P}({\rm CMB}) - {\rm Y}_{\rm P}({\rm SBBN,WMAP}) = 0.077 \pm 0.075.
\eeq
These differences are consistent with zero at $\sim 1.0\sigma$.  
Within its currently large uncertainty, the CMB provides an independent 
measurement of \Yp~in the high redshift, prestellar Universe consistent 
with the SBBN and non-SBBN predicted primordial \4he abundances.

\section{Primordial Helium-4 From Extragalactic \hii~Regions}

Historically, the primordial \4he mass fraction has been determined 
from observations of helium and hydrogen recombination lines in 
low-metallicity, extragalactic \hii~regions such as Blue Compact Galaxies 
(BCG)~\cite{pagel,it98,it04,os,fk,plp}.  As the data set has become larger 
and more accurate, it has become clear that, at present, the systematic 
uncertainties in converting the recombination line intensities to helium 
abundances dominate over the statistical errors.  Very recently, \citet{it10} 
and \citet{aos10} have revisited the BCG data, paying special attention to 
the systematic errors.  From a linear extrapolation to zero metallicity of 
the helium and oxygen abundances derived from 96 spectra in 86 \hii~regions, 
\citet{it10} find \Yp(IT10) $= 0.2565 \pm 0.0010({\rm stat}) \pm 0.0050({\rm 
syst})$.  In contrast, \citet{aos10} concentrate on the spectra from only 
nine, highly selected BCGs.  A linear extrapolation of the helium and 
oxygen data for their 9 BCGs leads \citet{aos10} to \Yp(AOS10) $= 0.2528 
\pm 0.0028$, where the uncertainty is the error in the mean which, since 
systematic errors dominate, may be an underestimate.  Within the errors, 
\citet{it10} and \citet{aos10} are in agreement.  For comparison with the 
helium abundance predictions and the CMB-determined value discussed 
above, the central value of the BCG-inferred primordial mass fraction 
from \citet{it10} is adopted here, and their statistical and systematic errors 
are combined {\bf linearly}, leading to \Yp(IT10) $= 0.2565 \pm 0.0060$.  
Although this estimate of the primordial \4he abundance is larger than the 
SBBN-predicted values (see \S\ref{SBBND} and \S\ref{SBBNW}), the 
differences are consistent with zero at $\sim 1.3 - 1.4 \sigma$.  In contrast, 
the \citet{it10} helium mass fraction is smaller than the non-SBBN value 
predicted for the WMAP (and LSS) determined baryon abundance and 
\Deln, but only by $\sim 0.6 - 0.7\sigma$.  The CMB-determined primordial 
helium abundance is higher than the \citet{it10} value but, given the large 
uncertainty, the difference is consistent with zero at $\sim 0.9\sigma$.

\section{Summary And Outlook}

For SBBN (\nnu~= 3), using the observationally-inferred primordial deuterium
abundance or the WMAP data to constrain the universal density of baryons, 
the predicted primordial helium mass fraction is \Yp(SBBN) $= 0.2482 - 
0.2488$, with a $\sim 2 - 3\%$ uncertainty.  These estimates agree, within 
$\la 1.4\sigma$, with the primordial value inferred from observations of low 
metallicity extragalactic \hii~regions~\cite{it10,aos10}, \Yp(\hii) $\approx 
0.2528 - 0.2565~(\pm \approx 0.0060)$.  However, the CMB and LSS 
data provide some evidence in support of a non-standard cosmology with 
\nnu~$\neq 3$~\cite{wmap7}.  For the WMAP estimates of $\eta_{\rm B}$ 
and of \Deln, \Yp(BBN,WMAP)~$= 0.2639 - 0.2649$, with an uncertainty 
ranging from 0.0088 to 0.0108.  Although higher than the SBBN estimates 
of \Yp, as well as those inferred from direct observations \cite{it10,aos10}, 
within the relatively large errors, they are consistent with them.  By 
combining the WMAP temperature anisotropy power spectrum with data from 
other CMB experiments \cite{acbar,quad}, \citet{wmap7} have presented evidence 
for an independent, high redshift, prestellar detection of helium, 
\Yp(CMB)~$= 0.326 \pm 0.075$.  Within its very large errors, this value 
too, is consistent with the others reviewed here.

It is anticipated that data from the Planck experiment \cite{planck} will 
result in significant reductions in the uncertainties in $\eta_{\rm B}$ 
and \Deln, as well as in the CMB estimate of \Yp.  According to \citet{hamann}, 
Planck will reduce the uncertainty in the baryon abundance parameter by 
more than a factor of two, from a WMAP value of $\sigma(\eta_{10}) \approx 
0.145$ to $\sigma(\eta_{10}) \approx 0.063$ and, will reduce the uncertainty 
in \Deln~by a factor of $\sim 3$, from $\sigma($\Deln)~$ \approx 0.76 - 0.88$, 
to $\sigma($\Deln)~$ \approx 0.26$.  If Planck achieves these reductions, the 
uncertainty in the BBN-CMB predicted helium abundance will be reduced by 
a factor of $\sim 3$, to $\sigma($\Yp)~$ \approx 0.0034$, resulting in a more 
accurate determination of \Yp~than is currently available from the direct 
observations of helium in extragalactic \hii~regions.  

The Planck experiment will also have improved sensitivity to a direct 
detection of primordial helium \cite{ichikawa}.  According to \citet{hamann} 
and \citet{ichikawa}, the uncertainty in the CMB value of \Yp~will be reduced 
by a factor of $\sim 5 - 7$, from $\sigma($\Yp)~$ \approx 0.075$ \cite{wmap7}, 
to $\sigma($\Yp)~$ \approx 0.011 - 0.014$.  Although still large, this 
uncertainty should be small enough that truly primordial helium will 
be discovered at more than the $5\sigma$ confidence level.
 
\acknowledgments
The research of GS is supported at The Ohio State University 
by a grant from the US department of Energy.  I am pleased to 
acknowledge informative correspondence with E. Komatsu. 

{}

\end{document}